\documentclass[twocolumn]{aastex631}

\newcommand\rev[1]{\textbf{#1}}

\graphicspath{{./}{Figures/}}

\begin{document}

\title{Pits on Jupiter Family Comets and the age of cometary surfaces}

\author[0000-0003-2354-0766]{Aurélie Guilbert-Lepoutre}
\affiliation{LGL TPE, UMR 5276 CNRS, Universit\'{e} Lyon 1, ENS, Villeurbanne, France}

\author[0009-0006-6704-9388]{Selma Benseguane}
\affiliation{LGL TPE, UMR 5276 CNRS, Universit\'{e} Lyon 1, ENS, Villeurbanne, France}

\author{Laurine Martinien}
\affiliation{LGL TPE, UMR 5276 CNRS, Universit\'{e} Lyon 1, ENS, Villeurbanne, France}

\author[0000-0001-9082-4457]{Jérémie Lasue}
\affiliation{IRAP, UMR 5277 CNRS, Université de Toulouse 3, CNES, Toulouse, France} 

\author{Sébastien Besse}
\affiliation{European Space Agency (ESA), European Space Astronomy Centre (ESAC), Villanueva de la Cañada, Spain}

\author{Bj\"orn Grieger}
\affiliation{Aurora Technology B.V. for the European Space Agency, ESAC, Madrid, Spain}

\author[0000-0002-5644-2022]{Arnaud Beth}
\affiliation{Department of Physics, Imperial College London, Prince Consort Road, London SW7 2AZ, UK}

\begin{abstract}
Large and deep depressions, also known as pits, are observed at the surface of all Jupiter Family Comets (JFCs) imaged by spacecraft missions. They offer the opportunity to glimpse into subsurface characteristics of comet nuclei, and study the complex interplay between surface structures and cometary activity. 
This work investigates the evolution of pits at the surface of 81P/Wild~2, 9P/Tempel~1 and 103P/Hartley~2, in continuation of the work by \citet{benseguane2022}, on 67P/Churyumov-Gerasimenko. 
Pits are selected across the surface of each nucleus, and high-resolution shape models are used to compute the energy they receive. A thermal evolution model is applied to constrain how cometary activity sustained under current illumination conditions could modify them. 
Similarly to what was found for 67P, we show erosion resulting from water-driven activity is primarily controlled by seasonal patterns, unique to each comet as a consequence of their shape and rotational properties. However, progressive erosion sustained after multiple perihelion passages is not able to carve any of the observed pits. Instead, cometary activity tends to erase sharp morphological features: they become wider and shallower over time.
Our results reinforce the evolutionary sequence evidenced from independent measurables to transform ``young'' cometary surfaces, with sharp surface topography prone to outbursts, into ``old'' cometary surfaces. 
Finally, we suggest that the mechanism at the origin of pits on JFCs should be able to carve these structures in a region of the solar system where water ice does not sublimate: the Centaur phase thus appears critical to understand JFCs surface properties.
\end{abstract}



\section{Introduction}\label{sec:intro}

The surfaces of comet nuclei display a diversity of morphological features \citep[terraces, fractures, or boulders for example,][]{massironi2015, el-maarry2019} that have a complex interplay with cometary activity and the cycle of material across the nucleus \citep[e.g.][for a review]{pajola2022}. 
Surface depressions offer the opportunity to connect the subsurface properties of comet nuclei with thermophysical processes actively shaping them \citep[e.g.][]{vincent2015a, davidsson2022b, benseguane2022}.
Shallow depressions (a few meters deep) generally observed on smooth terrains might be seasonal in nature, shaped by sublimation-driven activity orbit after orbit \citep{groussin2015, vincent2016, el-maarry2017, birch2019, bouquety2021a, davidsson2022b}. 
However, larger depressions, typically tens to several hundred of meters deep, are also observed \citep{vincent2015, el-maarry2019}, and cannot be linked with such seasonal activity \citep{benseguane2022}. 
These structures, also known as pits, have been observed on all the Jupiter-family comets (JFCs) imaged by spacecraft \citep{pajola2022}. 
At the surface of 67P/Churyumov-Gerasimenko (67P hereafter), they are mostly present on the northern hemisphere \citep{leon-dasi2021}. We recently studied their evolution through erosion, as a result of thermally-driven water ice sublimation under current illumination conditions \citep{benseguane2022}. We showed that through sustained cometary activity, erosion tends to erase such sharp morphological features. 
This result has implications for the origin of pits, 
as the modeled pits evolution suggests that none of the structures observed at the surface of 67P could be formed through progressive erosion on a typical JFC orbit.
This needs to be consolidated, or refuted, through the study of the evolution of similar structures seen on other comets.

\section{Characteristics of the pits observed on spacecraft targets}

\subsection{19P/Borrelly}
In 2001, the \emph{Deep Space~1} explored comet 19P/Borrelly, revealing an elongated and extremely dark nucleus \citep{soderblom2002}, with a highly variegated surface that can be divided into two terrain units: smooth and mottled terrain. A number of rounded depressions which could be identified as pits are visible in the mottled terrain, down to a scale of $\sim$200~m. These pits have a similar size, which supports the fact that they might not be related to impact craters, but rather the result of activity-related mechanisms \citep{vincent2015}.
While the smooth terrain is located in the sunward direction, the mottled terrain appears largely inactive as it is not associated with sources of gas and dust. 
The best spatial resolution obtained from \emph{Deep Space~1} is of the order of 50~m per pixel, compared to the average 10 to 15~m per pixel obtained by the subsequent  flybys missions described below. As a consequence, it is not possible to exploit the terrain model derived from the flyby images, and sample pits in any useful way for this study.

\subsection{81P/Wild~2}
The \emph{Stardust} images have revealed the presence of pits on the surface of 81P/Wild~2 (81P hereafter). They vary in both size and shape, some structures reaching several dozens of meters and up to $\sim$2~km in diameter.
\citet{brownlee2004} identified two types of pits: circular and irregular-shaped. Circular pits further exhibit two primary morphologies: pit-halo and flat-floored. 
Their origin has been presumed to be linked with impacts, possibly combined with sublimation and ablation processes. Indeed, hypervelocity impact experiments have successfully replicated pit-halo and flat-floor craters by impacting resin-coated sand with different degrees of porosity \citep{brownlee2004}. 
In that framework, the lack of small impact structures (with sizes $<$0.5~km) would be attributed to surface erosion, or a limited number of impactors within the corresponding size range.
Additionally, 81P exhibits non-circular depressions which have been assumed to be formed by a combination of sublimation, mass wasting, or ablation processes \citep{brownlee2004}.

\subsection{9P/Tempel~1}
Observations by \emph{Deep Impact} and \emph{Stardust/NExT} revealed that 9P/Tempel~1 (9P hereafter) has a very pitted surface \citep{belton2013}, with 380 pits ranging in diameter from tens to hundreds of meters (up to $\sim$900~m) and a depth of up to 25~m. Two of these depressions are considered as plausible impact craters \citep{thomas2007}. \citet{belton2013} inferred that JFCs would enter the inner solar system lacking ``primitive'' craters (i.e. formed through an intense, early collisional bombardment), and that most of these pits would likely result from outbursts of cometary activity. 
Indeed, they suggested that outbursts could account for the formation of 96\% of them, and the process could contribute to a significant portion of total nucleus mass loss, in addition to sublimation. 
Finally, \citet{belton2013} proposed that a few acute depressions may have resulted from sinkhole collapse, because the expected formation timescale for these surface structures substantially exceeds the corresponding sublimation timescale.

\subsection{103P/Hartley~2} 
Similarly to 9P, the surface of comet 103P/Hartley~2 (103P hereafter) displays depressions indicative of a formation process different than impacts. \citet{brucksyal2013} proposed that most surface structures, including circular depressions or pits, could be the products of evolving jets arising from vents, active during several orbits. 
In this framework, surface material located on the periphery of a vent could fall into pits or cracks during periods of low activity, eventually leading to shallower structures. This would bring warmer material in contact with the colder, icy material located at the bottom of the vent. This process could also apply to material tumbling from scarps and ridges, both at the surface of 103P and 9P \citep{farnham2013b}. 
The relation between jets and pits was supported by \citet{thomas2013b}, who additionally investigated the hypothesis of collapsing subsurface cavities.

\subsection{Ensemble properties}\label{ensemble}
Taken altogether, these observations suggest that pits may be ubiquitous on cometary surfaces, and that a link with cometary activity may exist. 
Moreover, pits observed on 9P, 81P, 103P and 67P display some morphological similarities in shape and dimensions \citep{vincent2015, ip2016}. 
However, some of the pits observed on these comet nuclei exhibit a lower depth-to-diameter (d/D) ratio compared to 67P. 
Active pits on 67P have an average d/D of $\sim$0.73, while inactive pits have a shallower d/D of $\sim$0.26 \citep{vincent2015}.
The observed average d/D is $\sim$0.2 for 81P \citep{kirk2005,vincent2015}, and $\sim$0.1 for 9P \citep{thomas2013a}.
Also, the aspect of comet 103P's surface resembles that of 9P, exhibiting a smooth appearance with no evident deep pits \citep{ip2016}.

~

In this work, we want to understand how cometary activity may modify these surface structures, and whether signatures of their formation process can be inferred from their expected evolution through sustained activity. 
We thus apply the same method as used to study the evolution of pits at the surface of 67P \citep{benseguane2022} to quantify the amount of erosion sustained by pits at the surface of 9P, 81P and 103P under their current illumination conditions.
We note that 19P/Borrelly cannot be included in this study as there is no shape model of its nucleus with a sufficient spatial resolution to be used.
We are interested in two quantities: the erosion sustained during each orbital revolution (i.e. the erosion per orbit in the following), and the erosion sustained as a result of multiple revolutions.
We summarize in Section \ref{sec:methods} the different steps of our method. Results for each comet are presented in Section \ref{sec:results} and discussed in Section \ref{sec:discussion}.


\section{Methods}\label{sec:methods}

\subsection{Shape models for comets 81P, 9P and 103P}
For each comet we study, a high resolution shape model is key to apply our surface energy model. We thus use the highest spatial resolution available for each nucleus, so to capture the effects of both their global shape, and local topography, pits in particular: 
\begin{itemize}
    \item[] 81P -- derived from the \emph{Stardust} Navcam images by \citet{farnham2005}\footnote{https://pdssbn.astro.umd.edu/holdings/sdu-c-navcam-5-wild2-shape-model-v2.1/dataset.shtml},
    \item[] 9P -- derived from the images obtained by the \emph{Deep Impact} and \emph{Stardust} missions by \citet{farnham2013} \footnote{https://pdssbn.astro.umd.edu/holdings/dif-c-hriv\_its\_mri-5-tempel1-shape-v2.0/dataset.shtml},
    \item[] 103P -- derived from images obtained by \emph{EPOXI} mission by \citet{farnham2013a}\footnote{https://pdssbn.astro.umd.edu/holdings/dif-c-hriv\_mri-5-hartley2-shape-v1.0/dataset.shtml}. 
\end{itemize}

Shape models for comets 81P, 9P, and 103P are not as spatially-resolved as those available for comet 67P. This lower quality (for our purpose) is to be expected, as these were derived from limited observations during flyby missions, whereas 67P was escorted during the 2~years of an orbiting mission. Again, we note that we had to exclude 19P from our study, as no available shape model has a good-enough resolution for our study.
More specifically, the shape models of 103P and 9P do not always reach the spatial resolution to unambiguously show features as deep as the pits observed on corresponding surface images. 
For example, the southern hemisphere of 81P was not observed during the \emph{Stardust} flyby: missing data has been completed by a smooth ellipsoid of revolution based on the average surface ellipsoid of the comet. 

It is important to mention that the smoother appearance of depressions on the surfaces of the studied comets is not solely due to the comparatively lower resolution of their shape models relative to 67P, but also because some of the pits observed on these comets exhibit a lower depth-to-diameter (d/D) ratio compared to 67P, as described in section \ref{ensemble}. 
Despite these limitations, we are able to identify a number of pits and alcoves on the shape models. We select facets similarly to those of 67P, located on the plateaus, walls and bottoms of each structure, to the best of our ability and the local resolution of these shape models.

\subsection{Selection of pits on each comet}

We select a minimum of 10 pits of each nucleus, located across all latitudes of each nucleus. Indeed, we showed that seasonal effects are dominating the global erosion trends on 67P \citep{benseguane2022}, so covering the entire nucleus is important in this study too.
We thus apply selection criteria similar to 67P. First, we aim to sample latitudes as much as possible, to assess the influence of seasonal mechanisms. 
Second we focus on large pits (rather than small and shallow depressions as seen on smooth terrains of 67P), characterized by steep walls and flat bottoms, and have sizes ranging from tens to hundreds of meters. 
Indeed, pits ranging from $\sim$150~m to $\sim$1~km do exhibit a size-frequency distribution that is similar for 67P and those observed on 9P and 81P \citep{vincent2015, ip2016}.
Effectively, we exclude smaller thermokarst features from this study \citep{bouquety2022}, as we did for 67P, because their formation and evolution appears to be different than the large and deep circular pits \citep[see][and references therein]{benseguane2022}.
On the corresponding shape models, we select multiple facets on different sides of each pit (plateaus, bottom and walls). 
When we select a pit on images that is not easily identified on the corresponding shape models, we pick facets with appropriate latitude and longitude. 
In these limited cases, our modeling outputs would serve the purpose of constraining the seasonal trends, and explore possible evolution scenarios for the corresponding surface features. 
For that same purpose, though we mostly select facets located in the northern hemisphere of 81P, we also select a group of facets in its southern hemisphere, even if there is no direct evidence for pits. 
This will allow to compare erosion rates across the entire nucleus to form a more complete picture.
For that purpose, we not only select several pits across the entire surface of 103P, but also add single facets randomly distributed across the surface in order to test the effects of this nucleus' unique elongated shape and complex rotational properties.
For each facet of the shape models, we then compute the thermal environment, including self-heating and shadowing, either by neighboring facets or due to the complex global morphology of the nucleus as described below.

\subsection{Surface energy and thermal evolution model}

    
  
   
    

The surface energy and thermal evolution models used in this study are described in \citet{benseguane2022}. We provide below a brief summary.
The total energy $\mathcal{E}$ received by each facet selected for this study is the sum of different contributions: direct insolation $E_{\odot}$ (accounting for shadowing effects) and self-heating, i.e. the energy received by reflection and emission from neighboring facets in the visible $E_\mathrm{VIS}$ and infrared $E_\mathrm{IR}$. Each contribution is given below.
Direct insolation is given by:
\begin{equation}
    E_{\odot} = \frac{F_{\odot}}{r_H^2}~ \cos{\xi}
\end{equation}
\begin{equation}
E_\mathrm{VIS} = \sum_T ~~ \mathcal{A}_T~ \frac{F_{\odot}}{r_H^2} \cos{\xi_{T}} ~\frac{S_T}{\pi} ~\frac{\cos{\zeta_{T}}~\cos{\zeta_{R}}}{\delta^2_T} 
\label{e_vis}
\end{equation}
\begin{equation}
E_\mathrm{IR} = \sum_T~ \varepsilon\sigma T_T^4~ \frac{S_T}{\pi}~~\frac{\cos{\zeta_{T}} \cos{\zeta_{R}}}{\delta_T^2 } 
\label{e_ir}
\end{equation}
with $F_{\odot}$~[W~m$^{-2}$] the solar flux at 1~au, $\mathcal{A}_T$ the Bond albedo of an emitting facet ($\mathcal{A}_T$=0.06 for all calculations), $\xi_T$ its local zenith angle, $S_T$ its surface, $\zeta_{T}$ the angle between the normal of the transmitter and the receiver facets, $\zeta_{R}$ the angle between the normal of the receiving and the emitting facets, and $\delta_T$ the distance between the two facets. The value of the emissivity $\varepsilon$ is 0.95 for all calculations.
$T_T$ is computed by considering direct insolation only, without any prerequisite knowledge of the importance of the self-heating contributions. When an emitting facet experiences night during a given timestep, we set a minimum threshold of $T_T$~=~20~K. The heliocentric distance $r_H$~[au] and the local zenith angle $\xi$ both vary with time, as described in the following subsection.

The total energy $\mathcal{E}$ is used in the surface boundary condition of a 1D thermal evolution model. This condition is given by:
\begin{equation}
        (1-\mathcal{A}_R)~ \mathcal{E}= \varepsilon \sigma T^4 + \kappa \frac{\partial T}{\partial r} + f_{H_2O}~ \Delta H_{H_2O}~ Q_{H_2O}
        \label{surf_bound}
\end{equation}
where $\mathcal{A}_R$ is the Bond albedo (with a value of 0.06 in all calculations) of the facet for which we compute the energy balance, $\mathcal{E}=E_{\odot}+E_\mathrm{VIS}+E_\mathrm{IR}$ the total energy received at its surface, $T$ [K] the surface equilibrium temperature, $f_{H_2O}$ the fraction of the facet covered by water ice, $\Delta H_{H_2O}$ the latent heat of water ice sublimation, and $Q_{H_2O}$ [kg~m$^{-2}$~s$^{-1}$] the corresponding sublimation rate. 
We aim at constraining how the patterns of energy received at the surface (diurnal but most significantly seasonal) influence the activity of each nucleus and the erosion of its surface features. Thus, we assume that thermal and physical characteristics are the same for each comet so that the contributions of varying parameters may be removed. 
In order to make our results comparable to our study of pits on 67P, we use the same set of initial parameters, a hypothesis with consequences discussed in Section \ref{sec:discussion}. For instance, we assume that the material is a simple mixture of two components, water ice and dust, with a mass fraction ratio of 1 and a porosity of 75\%. The resulting thermal conductivity is reduced by a Hertz factor of 0.005 to account for the limited contact between grains in this porous structure \citep[e.g.][for further description of this parameter]{guilbert-lepoutre2023}.

\citet{benseguane2022} showed the influence of each of these parameters on the erosion rates sustained on 67P, so we do not repeat these here since the effects are the same. The chosen values for the initial parameters are listed and justified in \citet{benseguane2022}. Since the behavior of cometary material in our model depends on the energy received at the surface of each facet and its heating rate, we can extend the conclusions of \citet{benseguane2022} on the influence of these parameters for this work.
We note that including CO and CO$_2$ in the volatile mixture was not altering the evolution trends in any significant manner, since the most significant source of erosion was the sublimation of water ice. Therefore, we do not add these species in our mixture. 
The thermal evolution model includes usual features such as heat and gas diffusion, phase transitions for water ice (crystallization and sublimation), drag of dust particles by the vapor phase, and formation of a dust mantle at the surface \citep{lasue2008}. 



\subsection{Orbital considerations for each comet} 

The surface energy model is computed with a timestep of 8~minutes for each comet. This allows to achieve a good description of the diurnal patterns of heating and resulting activity, for any combination of spin state and shape. We note that the extreme members in that respect are 67P studied by \citet{benseguane2022}, and 103P which has a complex rotation state \citep{belton2013,knight2015}.
For each timestep, we first retrieve the coordinates of the subsolar point using SPICE kernels available on the WebGeocalc platform\footnote{https://wgc.jpl.nasa.gov:8443/webgeocalc/\#NewCalculation}$^,$\footnote{http://spice.esac.esa.int/webgeocalc/\#NewCalculation}, which contain the information on each nucleus's rotation state, pole orientation, and orbital parameters. Then, the insolation geometry for each facet is computed with respect to these subsolar point coordinates. As a result of 103P's complex rotation state, SPICE kernels are probably only valid for the duration of the \emph{EPOXI} flyby: extrapolations before and after the flyby duration may not be accurate: the impact of this complex spin state will be discussed with the results. 
We run our thermal evolution calculations to study the impact of current illumination conditions. However, each comet has ``acquired'' its current orbit following a distinct orbital evolution, and has been holding it for a different period of time since its latest orbital change. 
We thus use a different number $x$ of orbital revolutions for each comet, to reflect their recent past history, based on backward dynamical integrations performed by \citet{ip2016}: $x=$~6 orbits for 81P, $x=$~13 orbits for 9P, and $x=$~20 orbits for 103P. 
In the following, we present both the erosion sustained during each orbital revolution which allows to compare comets, as well as the total erosion sustained under current illumination conditions after $x$ revolutions which allows to assess whether erosion driven by the sublimation of water ice could carve pits as they are observed on each individual comet.


\section{Thermal processing of pits}\label{sec:results}

\subsection{81P/Wild~2}

\begin{figure}[ht]
    \centering
    \includegraphics[width=\columnwidth]{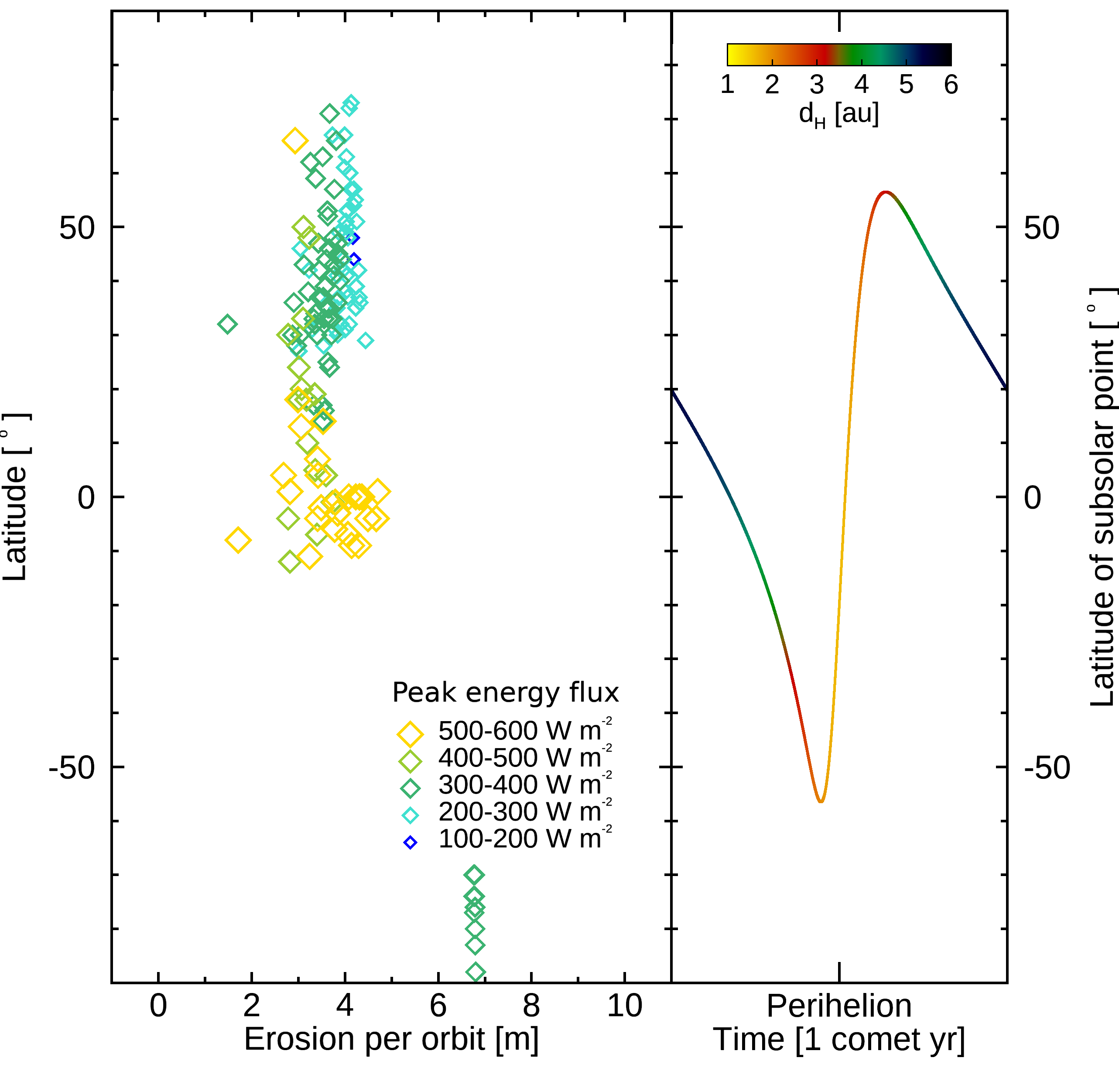}
    \caption{Left: erosion of each selected facet on the surface of 81P as a function of latitude and energy peak. Right: latitude of the subsolar point as a function of heliocentric distance.}
    \label{fig:lat_dh_81p}
\end{figure}

The spin state of 81P is such that the subsolar point crosses a large range of latitudes near perihelion (from -60$^{\circ}$ to 60$^{\circ}$, see Fig.~\ref{fig:lat_dh_81p}). As a result, the total amount of energy per orbit received by each facet selected in our study is relatively uniform across the surface. 
The slight asymmetry between the pre- and post-perihelion latitudes and corresponding heliocentric distances results in the southern hemisphere receiving on average almost twice as much energy as the northern hemisphere. 
As for 67P \citep{benseguane2022}, the latitudinal effects (i.e. seasonal) dominate the energy distribution at the surface of 81P, but the local shape can also play a key role.
Indeed, we note that only a few facets located on the walls of some pits receive lower amounts of energy in total, about half the maximum amount received by others, as a result of shadowing effects from neighboring facets.
The contribution of self-heating to the total energy is relatively small, but it can account for up to 30\% of the total energy in the shadowed regions where direct insolation is weak. It should be noted, however, that because 81P's pits are quite large \citep[$\sim$2~km for the largest one,][]{brownlee2004}, most facets are ultimately exposed to direct insolation.

The erosion resulting from the sublimation of water ice is, consequently, relatively uniform for all selected facets in the north (Fig.~\ref{fig:lat_dh_81p}), amounting to 4 to 5~m per orbital revolution, and $\sim$15 to 30~m at most after 6 orbital revolutions 
(Table~\ref{tab:recap}).
We note here that the southern hemisphere may erode more than in the northern hemisphere. Indeed facets we considered in the south sustain a maximum erosion of the  order of 40~m after 6 revolutions. Nonetheless, the lack of actual images and the corresponding poor quality of the shape model for this region of the nucleus does not allow us to draw any further conclusion. 
These results imply that erosion driven by water sublimation is not likely the primary process responsible for the formation of the large pits of diameters up to 2~km studied here.
We note that in general, pits on 81P are large enough for facets located at their bottom to behave similarly to facets located on the surrounding plateaus. Both ``parallel'' planes thus erode in the same way, so that the depth of these features should be expected to remain relatively constant with time.

\begin{table}[ht]
    \centering
    \caption{Summary of results, including data for 67P from \citet{benseguane2022}: number of orbital revolutions, perihelion distance $q$, maximum erosion, compared to the pits' diameter and depth. $\star$: for 81P the maximum is given for facets corresponding to observed regions of the surface, excluding the southern hemisphere.}
    \begin{tabular}{ccccccc}
    \hline
    Comet & Orbits & q & Max  & Pits & Pits\\
          &        &   & erosion & diameter & depth\\
    & [\#] & [au] & [m] & [m] & [m]\\
    \hline 
    67P & 10 & 1.24 & 77.23  & 100s       & 10s-100s\\
    81P           & 6  & 1.59 & 28.25$^{\star}$  & 100s-1000s & 10s-100s \\
    9P            & 13 & 1.54 & 83.81  & 10s-100s   & 10s\\
    103P          & 20 & 1.06 & 265.10 & 10s-100s    & 10s\\ 
    \hline
    \end{tabular}
    \label{tab:recap}
\end{table}

\subsection{9P/Tempel~1}



\begin{figure}[ht]
    \centering
    \includegraphics[width=\columnwidth]{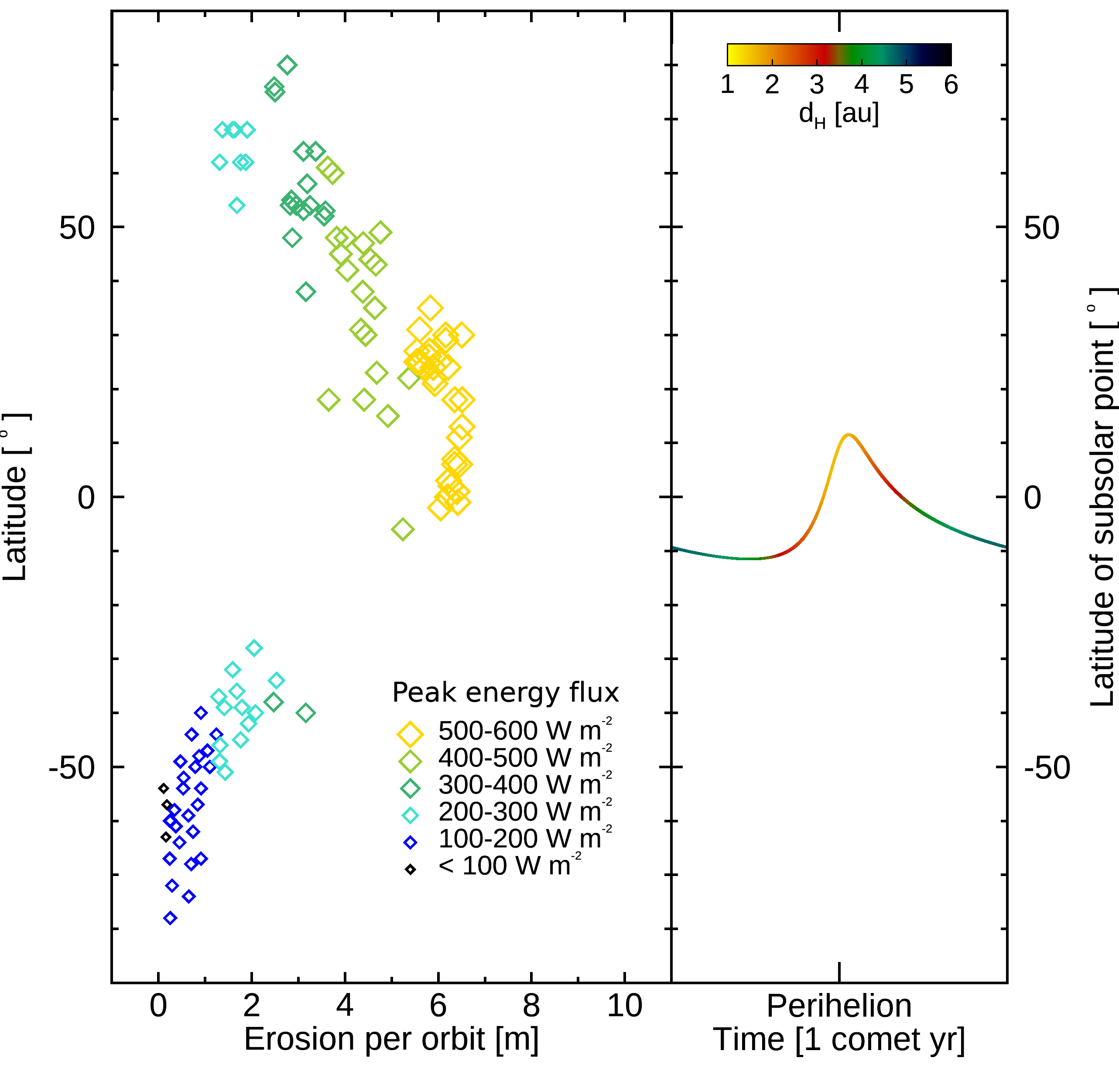}
    \caption{Left: erosion of each selected facet on the surface of 9P as a function of latitude and energy peak. Right: latitude of the subsolar point as a function of heliocentric distance.}
    \label{fig:lat_dh_9p}
\end{figure}

Comet 9P is, in the context of this study, the ``simplest'' comet. Due to its low obliquity, the largest amounts of energy are received at perihelion by facets located in the 0 to 20$^{\circ}$ latitude range. 
The northern hemisphere receives slightly more energy than the southern hemisphere, and both patterns of peak and total energy do correlate well with the latitude of the subsolar point (see Fig.~\ref{fig:lat_dh_9p}). In this regard, seasonal patterns resulting from a combination of shape and rotational properties also dominate the energy patterns, in a way even more obvious than for 67P or 81P. We note local differences for facets selected on each pit, which suggest patterns of differential erosion similar to those obtained for 67P \citep{benseguane2022}.
Erosion caused by water-driven outgassing is larger than for 81P, and because we consider a total of 13 orbits, it can eventually become substantial in equatorial regions. 
However, it never exceeds a hundred meters in total (Table~\ref{tab:recap}), whereas the observed dimension of pits can reach several hundred meters across for the largest ones \citep{thomas2013}.

\subsection{103P/Hartley~2}

Comet 103P is in non-principal axis rotation with apparently changing component rotation periods \citep{belton2013,knight2015}. Its rotational properties are, in fact, so complex that the SPICE kernels have a limited range of validity around the \emph{EPOXI} flyby of the nucleus, and that propagating coordinates for the subsolar points to the whole orbit is not necessarily possible. This is nonetheless the best we can do at this point to assess the influence of activity on the evolution of surface features. We must keep this effect in mind to interpret our results.
The nucleus spins in an excited long-axis mode, with its rotational angular momentum per unit mass and rotational energy per unit mass slowly decreasing while the degree of excitation in the spin increases through perihelion passage \citep{belton2013}. To further complicate the picture, the nucleus has a very elongated shape. 

\begin{figure}[ht]
    \centering
    \includegraphics[width=\columnwidth]{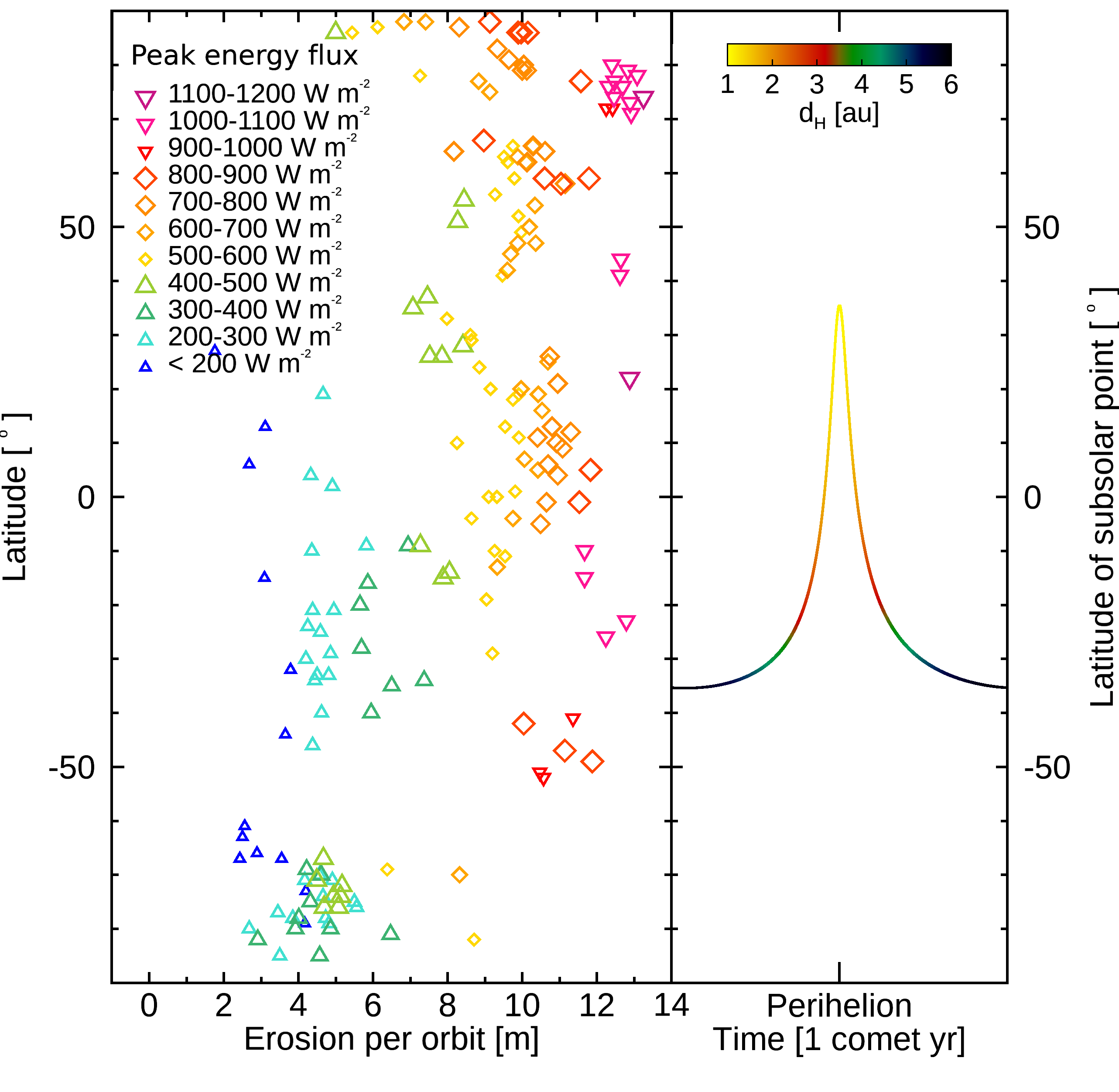}
    \caption{Left: erosion of each selected facet on the surface of 103P as a function of latitude and energy peak. Right: latitude of the subsolar point as a function of heliocentric distance.}
    \label{fig:lat_dh_103p}
\end{figure}

These characteristics are reflected in the complex distribution of energy received by the nucleus' surface (Fig.~\ref{fig:lat_dh_103p}), which exhibits not only a latitudinal trend (as observed for the other comet nuclei), but also strong variations across longitude, especially around the equator region. Overall, equatorial regions and nearby northern latitudes receive a substantial amount of energy around perihelion, while the southern and extreme western equatorial regions receive less energy during this period. 
The contribution of self-heating to the total energy on 103P is minimal, accounting for less than 10\% of the total energy. This contribution is extremely low compared to 67P or 81P. This is primarily due to pits on 103P having a low d/D (depth-to-diameter) ratio compared to 67P or 81P (see Table~\ref{tab:recap}). Additionally, the low spatial resolution of the shape model may play a role in limiting the ability to effectively reproduce shadowing and self-heating effects on a scale smaller than 10~m. The two effects might compensate each other, however we can calculate that with an additional 10\% of surface energy due to self-heating, the final erosion would be enhanced by $\sim$15\% for the most eroded facets.

Erosion on 103P is strongly correlated with the peak of energy, received at, or close to perihelion (see \ref{fig:lat_dh_103p}). However because of 103P's complex rotation, almost all facets are ultimately exposed at perihelion. This correlation can be attributed to the fact that erosion on 103P occurs predominantly during brief periods of intense heating, outside of which, the energy is insufficient to cause facets to erode. This is in contrast to 81P or 9P where energy is more consistently distributed throughout the entire near-perihelion passage, instead of occurring in brief peaks. At perihelion 103P is also the comet closest to the Sun, at a distance of 1.05~au, compared to 81P (1.59~au), 9P (1.54~au), or even 67P (1.24~au). As a result, 103P is the comet that sustains the most erosion after 20 revolutions under current illumination conditions (Table~\ref{tab:recap}).


\section{Discussion}\label{sec:discussion}

\subsection{Erosion and the evolution of pits}

\begin{figure}[ht!]
    \centering
    \includegraphics[width=\columnwidth]{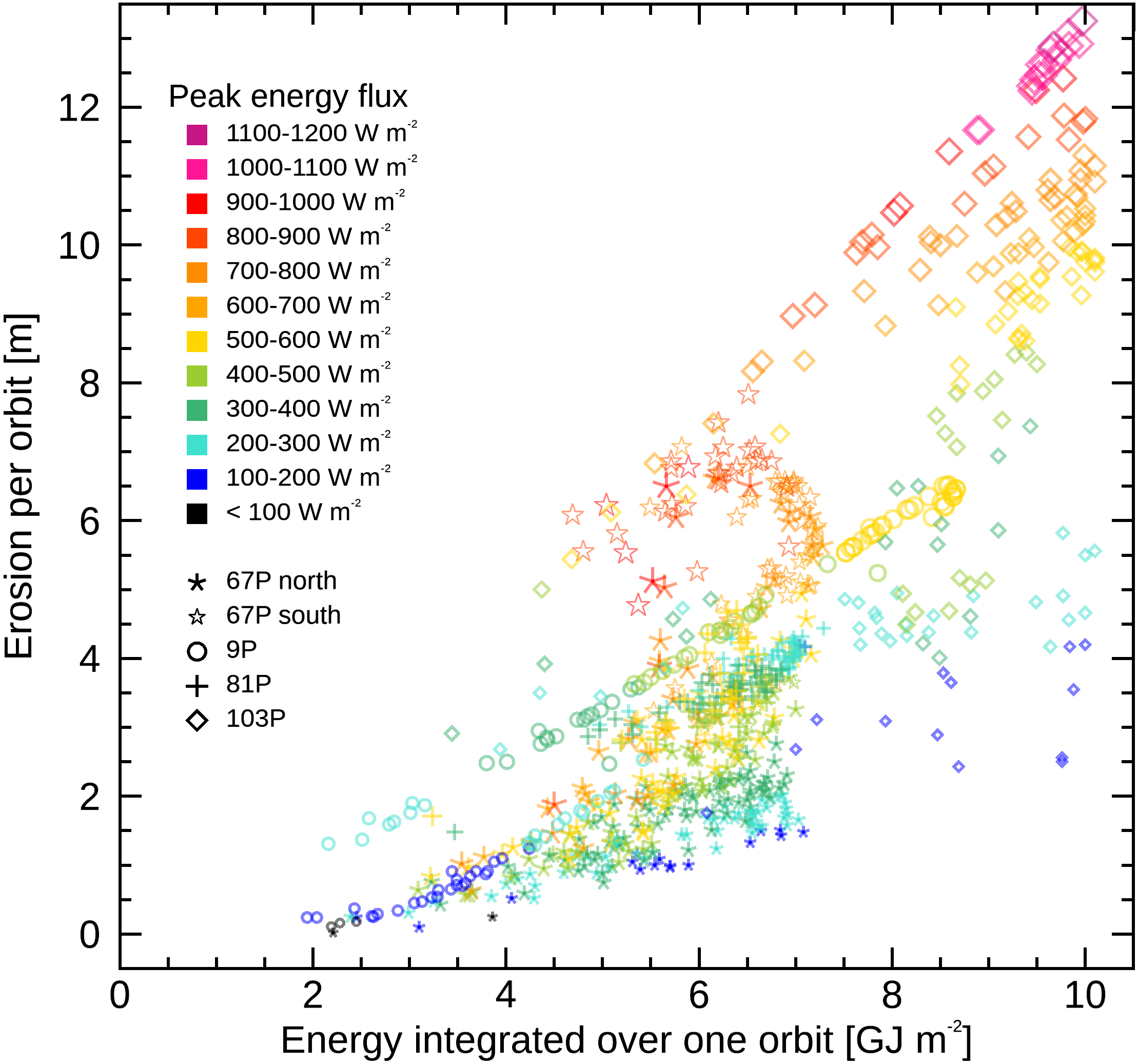}
    \caption{Erosion per orbit, as a function or total energy integrated over one orbit. The color code and the increasing size of symbols gives (increasing) the peak energy, usually received close to perihelion.} 
    \label{fig:energy}
\end{figure}

For all comets considered in this work, cometary activity driven by the sublimation of water ice, and the resulting erosion, are primarily controlled by direct insolation and thus display strong seasonal patterns. Indeed, the thermal processing of each facet considered in this study depends not only on the peak and total energy it receives, but also on how energy is delivered as a function of time, to produce subtle effects that appear unique to each comet nucleus, and in particular its shape and rotational properties, as shown in Figure~\ref{fig:energy}. 
For comets 9P, 103P, and 81P, no global shape effects are observed due to the lack of substantial shape irregularities on a global scale, unlike the accute shape of 67P. The unique shape of comet 67P, for instance, makes pits located near the neck of the nucleus more susceptible to shadowing effects by the smaller lobe \citet{benseguane2022}.
However, we found that local shape effects (i.e. linked to local topography) can be significant at the scale of a given pit. Self-heating can contribute with important fractions to the total energy in deep pits and steep cliffs of 81P for instance, accounting for ~30\% of total energy input. In contrast, it is minimal for 9P and 103P ($<$10\%), where surface features are wider and shallower.

Facets located at the bottom and on the wall of circular depressions can be affected by shadowing, compared to exposed plateau facets. Consequently, if these depressions are deep enough, they tend to become shallower over time due to water-driven erosion. This is due to the combination of two effects. First, plateaus tend to erode more than bottoms (being more exposed to direct insolation), so that pits tend to become shallower with time. In addition, walls sustain some differential erosion: over several perihelion passages, pits also become wider over time.
When this trend is not observed, it is because the corresponding pits are already large or shallow, i.e. not deep enough compared to their diameter, as seen in most of the pits on 9P and 103P, or pits are large enough that the bottom facets are directly exposed to the Sun, as in the case of the large pits of 81P. 
While 81P is the least eroded nucleus in our study, our results imply that most of its largest pits are already wide enough to prevent any further change in depth. We note here that additional processing of cliffs can occur following their collapses as observed for 67P \citep{vincent2016,pajola2017,el-maarry2019}. By this mechanism (not accounted for in our model), the filling of pits with debris material would tend to further reduce their depth.
Overall, we suggest that pits can reach a depth-to-diameter (d/D) that seems to prevent any further change due to erosion alone. 
Future studies could explore critical d/D thresholds in different illumination conditions, determining when pits are able to evolve or when their morphology becomes relatively fixed as a function of material properties. 
\citet{ip2016} suggested that the d/D ratios of the large pits are mostly within the range of 0.1–0.3. 
In comparison, active pits studied by \citet{vincent2015} have a large d/D ratio ($>$0.3) and a small diameter ($<$300~m). This statistical result coupled with backward dynamical integrations suggest that large circular depressions could have outgrown from the small and deep ones via erosive mass wasting of the surrounding areas \citep{cheng2013,vincent2015,vincent2016}.

Overall, we find that sharp depressions are likely erased with time as a result of sustained cometary activity. 
Most significantly, erosion sustained after the multiple perihelion passages is not able to carve large depressions with the observed size and shape, on any of the comets we studied. Of course, some limitations arise from our methodology, most notably from the assumed uniform thermal and physical characteristics, for all pits we have studied. Local heterogeneities (in composition, albedo or thermal properties for example) could actually enhance the local erosion computed in our simulations. 
Significant deviations from our results can only be achieved with extreme values for the initial parameters we have considered: for example, a combination of 70\% of surface water ice with a porosity larger of the order of 90\% can double the amount of erosion.
As a result, within the range of plausible parameters \citep[see][for a review]{benseguane2022}, erosion could be increased by up to 20-30\% at most: this does not affect our general trends, nor the general conclusion that cometary activity tends to erase sharp surface features.

\subsection{The case of 103P}
Comet 103P is an extreme example of the effects described above. 
In our simulations, the small northern lobe of the 103P's nucleus is very active and experiences the most erosion as a result of its preferential exposure to the Sun at perihelion. 
The northern lobe was indeed observed to be active during the \emph{EPOXI} flyby. More precisely, jets were clustered in the rough topography of the small northern lobe and mid- to northern part of the big lobe \citep{ahearn2011}.
Nevertheless, the observed high activity may have been enhanced by an abundance of volatile species in specific regions of the small lobe, not accounted for in our model. \citet{ahearn2011} indeed determined that different species were being ejected from the different parts of the nucleus, with H$_2$O vapor coming primarily from the waist and CO$_2$, H$_2$O ice, and organics coming primarily from the top of the small lobe. 
Taking into account a higher abundance of volatile species in the small lobe within our thermal evolution calculations results in an increased erosion compared to erosion yielded from a homogeneous nucleus assumption. 
This would further emphasize the contrast with the southern big lobe. \emph{EPOXI} additionally revealed distinct terrains on the nucleus, with a smooth ``waist'' connecting two rougher lobes \citep{ahearn2011,knight2013b,thomas2013}. No significant difference was noted in the concentration or appearance of pits between the two lobes though. In light of our results, we could argue that this might be caused by erosion being sustained in a similar manner across the two lobes.
We mentioned that due to 103P's complex rotation state, the kernels used to derive the latitude of the subsolar point might not be accurate outside of the \emph{EPOXI} flyby duration. Of interest to our results, the key effect of this spin state is primarily to expose the whole nucleus at perihelion, where peak energy fluxes are received at all latitudes to efficiently trigger water ice sublimation and erosion (Fig.\ref{fig:lat_dh_103p}). This occurs within the validity range of the SPICE kernels, so ultimately our results should not be severely affected.


~

Overall, we find that progressive erosion, driven by the sublimation of water ice under current illumination conditions, is not able to form pits at the surface of JFCs because: a) the total erosion, even after calculations taking into account several orbital revolutions, remains lower than the observed dimensions of pits, and b) it tends to erase sharp feature, which become shallower and wider with time.

\subsection{Implications for the aging of cometary surfaces}  
\begin{figure*}[ht]
    \centering
    \includegraphics[width=\textwidth]{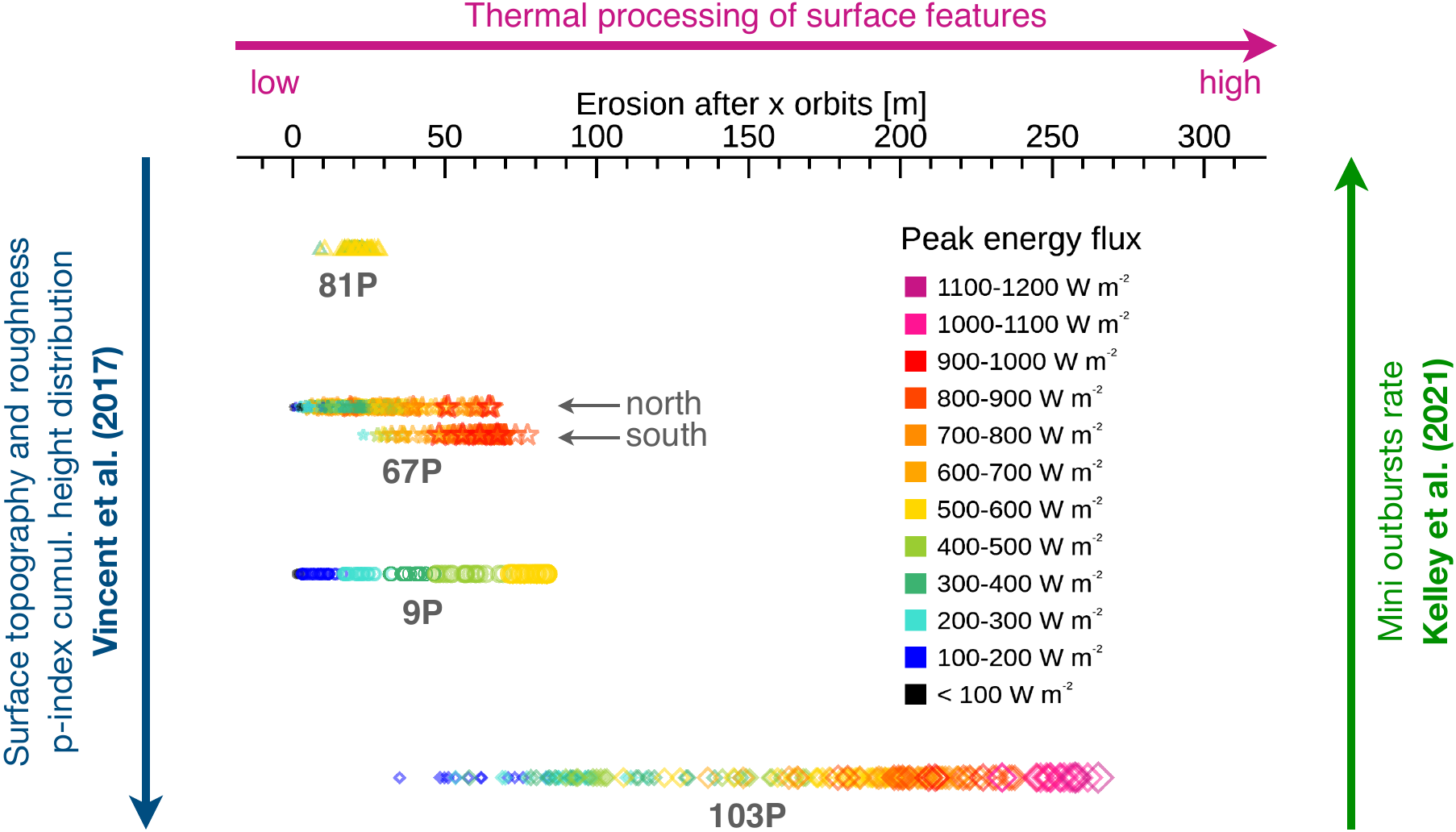}
    \caption{Illustration of the evolutionary sequence between young (81P) and old (103P) cometary surface. Total erosion is calculated for each facet after a given number of orbital revolutions unique to each comet nucleus (10 for 67P, 13 for 9P, 6 for 81P and 20 for 103P), to reflect their evolution under current illumination conditions. The color code (and the increasing symbol size) gives the (increasing) peak energy received by each facet, typically around perihelion. Sequences suggested from the cumulative distribution of surface roughness and topography by \citet{vincent2017a} and the rate of mini-outbursts by \citet[][excluding 81P]{kelley2021} are overlaid: the vertical spacing of comets on these corresponding scales is qualitative, to serve an illustration purpose.}
    \label{fig:sequence}
\end{figure*}

If sharp features are indeed erased by erosion, driven in our simulations by the sublimation of water ice, then as a corollary, we can infer that the deepest, most circular pits are likely the most primitive, or the best preserved pits. 
From the results of our thermal evolution model, including those obtained for 67P by \citet{benseguane2022}, we can ``rank'' the primitiveness of these surface structures observed on these four comet nuclei. Comet 81P would have the least processed pits (or best preserved), followed by the northern hemisphere of 67P, then its southern hemisphere. Comet 9P overlaps in the thermal processing space with 67P, with its southern hemisphere being relatively unprocessed compared to its equatorial region. Finally, 103P is by far the most thermally processed JFC we have studied, due to a combination of aggravating factors: smaller perihelion distance, larger number of orbital revolutions close to the Sun, a complex rotation that leads to the relatively uniform processing of the entire surface.

Interestingly, \citet{vincent2017a} had performed a statistical analysis of the distribution of large-scale topographic features on 67P, and found that cliff height correlates with surface erosion rates and follows a power law with an average cumulative power index of -1.69. They suggest that topography could be used to trace a comet's erosional history. In this framework, large and sharp cliffs would characterize primordial surfaces, while eroded surfaces would display smaller blocks (e.g. boulders, pebbles and dust). The power law index of the corresponding topography cumulative height distribution could indicate how primitive a comet nucleus is.
They performed the same statistical analysis of surface features observed on 81P, 9P and 103P \citep[see Table~2 of][]{vincent2017a}. They found that 67P and 81P would have experienced similar degrees of erosion, while comets such as 9P and 103P would be more eroded, in agreement with the suspected past dynamical histories for each comet \citep[see Figure~11 of][]{vincent2017a}.
They concluded that a comet recently entering the inner solar system would have a p-index of topographic height around -1.5. Older comets show larger power indices, up to about -2.3.

Based on these results, \citet{kokotanekova2018} proposed that the phase-function-albedo correlation they had previously found \citep{kokotanekova2017} might be explained by the erosion of pits and rough surface topography. Based on their hypothesis, rough surfaces with steep phase functions would gradually evolve toward smoother terrains with decreased phase function coefficients. 
Our results stem from a distinct method, providing a physical model to both empirical studies \citep{vincent2017a,kokotanekova2018}: they point to the same evolutionary sequence.
Assuming that this overall interpretation is correct, a decreasing phase function coefficient would provide a useful observable to characterize the level of erosion of a cometary surface.

With a very different prism, \citet{kelley2021} examined several outbursts observed on comet 46P/Wirtanen and found that mass estimates were similar to, or an order of magnitude larger than, the mini-outbursts observed at comets 9P and 67P. They hypothesized that mini-outbursts on comets could be associated with steep terrain features like cliffs and scarps, based on observations linking such mini-outbursts of comet 67P to these terrain features, and even their collapse \citep{vincent2019}. 
Based on this assumption, they analyzed the outburst frequencies of comets 67P, 9P, 46P, and 103P. They suggested that the observed differences may be related to distinct surface terrains. Comets 67P and 9P displayed significantly higher outburst frequencies compared to 46P and 103P. Interestingly, comet 46P would appear as an evolutionary intermediate between 103P (very processed) and 9P (moderately processed) in terms of surface topography and \rev{erosion implied from the work presented here}. Indeed, this comet has performed a number of orbital revolutions since its discovery similar to 9P, with a perihelion distance decreasing from $\sim$1.6~au (i.e. similar to 9P) to $\sim$1.05~au \citep[i.e. similar to 103P][]{krolikowska1996}. This is in agreement with the processing sequence proposed by \citet{vincent2017a} and \citet{kokotanekova2018}.

The results we present here agree with the evolutionary sequence proposed from independent observables \citep{vincent2017a, kokotanekova2018, kelley2021}.
With a distinct method, we can provide a physical framework to this evolutionary sequence that transforms ``young'' cometary surfaces, with sharp surface topography prone to spark mini-outbursts, into ``old'' cometary surfaces that are eroded and do not experience as many mini-outbursts, as summarized in Figure~\ref{fig:sequence}. By ``young'', what we mean here is that a comet nucleus' surface has undergone relatively little modification resulting from water driven activity, although the nucleus is located in a region where water is efficiently sublimating (typically with q$<$2.5~au). This is a consequence of each nucleus' unique past dynamical history and rotational properties. 
In this study we have only skimmed over the most recent influence of each comet's past orbital evolution by accounting for a number of perihelion passages, which amounts to $\sim$40 to 130~years of thermal processing at most (for 81P and 103P respectively). However the dynamical evolution since comet nuclei left the outer solar system reservoirs to reach the orbit on which they are currently observed is much longer, complex, and entails some thermal processing that is not accounted for in our work \citep[e.g.][]{gkotsinas2022}.
On the opposite, ``old'' refers to a surface that has significantly changed as a result of thermal processing, leading to water-driven activity and substantial surface erosion.

\subsection{On the origin of pits}

Various scenarios have been proposed in the literature to explain the origin of pits \citep[see][for a review on 67P, and the introduction of this study]{benseguane2022}. \citet{ip2016} found that such features with steep walls and flat bottoms, with sizes between 150~m and 1~km on 67P, have the same size frequency distribution as those on 81P and 9P. Therefore, it seems reasonable to assume that pits observed at the surface of JFCs share a similar origin, so that their morphological characteristics appear similar. These steep walls and flat bottoms make them different from the bowl-shape impact craters found on the Moon or asteroids \citep{brownlee2004, ip2016}. Although we cannot exclude that some pits may remain associated to impact events, these could rather be considered as a signature of some process related to cometary activity rather than the result of collisions. 

In light of the discussion above, we would like to highlight the conclusions of \citet{belton2013} who suggested that most pits on the surface of 9P surface would likely be the most common surface features related to outbursts of activity. Additionally, \citet{pozuelos2014} suggested that cometary outbursts could be at the origin of pits observed on 81P. From our results, key aspects of how water-driven cometary activity fuels the evolution of pits need to be recalled. First, we see that sharp features tend to be erased, as they become wider and shallower with time. Second, latitudinal effects are so strong that patterns of differential erosion tend to elongate initially circular features. If we assume that pits formed as cylindrical structures, erosion with time would lead to elongated features departing from this initial morphology.
In order to carve deep, almost circular structures, it is therefore crucial that water ice does not sublimate whenever pits are formed. Therefore, it appears that pits might have been formed before JFC nuclei cross the water snow line, a suggestion made by \citet{ip2016} who studied the past dynamical history of these comets.

The morphological characteristics of the least processed pits we could identify imply a formation scenario where a rather explosive mechanism, able to carve a large amount of cometary material in a short period of time, occurs in a region of the solar system where water ice is not sublimating, or the freshly-formed features would become progressively elongated and eroded.  
The Centaur phase experienced by each JFC \citep[e.g.][]{gkotsinas2022} may be key here to understand the origin of such surface features. In the giant planet region, several phase transitions do occur that can lead to cometary activity amongst Centaurs \citep[e.g. crystallization of amorphous water ice or CO$_2$ sublimation or segregation,][respectively]{guilbert-lepoutre2012, davidsson2021a}. Sudden thermally-induced events such as clathrate destabilization and the crystallization of amorphous ice could lead to outbursts of activity \citep{miles2016,wierzchos2020}, potentially leading to the formation of pits. Furthermore, several Centaurs are prone to recurrent, sporadic outbursts of activity, like 29P/Schwassmann-Wachmann \citep{wierzchos2020, clements2021, lin2023, betzler2023}, or 174P/Echeclus \citep{rousselot2016, kareta2019, rousselot2021}.

\subsection{Perspectives}

Our hypothesis for the formation of pits could be tested in the future by the \emph{Comet Interceptor} mission \citep{snodgrass2019}. This mission is designed to encounter a Dynamically New Comet, which typically experiences only limited processing in the giant planet region in comparison to JFCs. 
Observing no pits could imply that their origin is linked to a process that exclusively affects JFCs. Since the key difference between these two populations rests mainly on their orbital evolution, the formation mechanism of pits should be sought there, and the thermal processing entailed by JFCs' dynamical evolution.
Alternatively, a limited number of pits could be observed, which we might attribute to thermal processing prior to the flyby, possibly on the inbound part of the orbit, or the early processing prior to the ejection of the nucleus in the Oort Cloud. Comparisons with the characteristics of pits observed on JFCs (depth, diameter, location with respect to the subsolar point for example) would help to pin point the origin of these surface features. 
Finally, observing as many pits and sharp topography at the surface of such a pristine comet nucleus as on JFCs would suggest that these are signatures of mechanisms at play during the earliest stages of comet formation, rather than the signature of processes at play during the Centaur phase of JFCs. Indeed, the implication would be that such rough surface topography would be common to all comet nuclei before water sublimation sets in, regardless of their subsequent orbital evolution.

Evidence for the evolutionary sequence provided by this work and the prior studies by \citet{vincent2017a}, \citet{kokotanekova2018} and \citet{kelley2021} highlights the importance of more space- and ground-based observations of comet nuclei. In particular, the best way to verify the validity of this sequence is a) to increase the number of comets, and b) use multiple independent techniques to cross-check the resulting sequences. We advocate that programs targeting JFCs at various stages of their evolution will be of primary importance in advancing our understanding of these objects. In particular, space missions toward active or outbursting Centaurs (such as 29P/Schwassman-Wachmann~1) would certainly prove instructive. Missions toward and targeted telescope observations of less evolved Centaurs, especially those currently orbiting beyond Saturn and are less processed from a statistical point of view \citep{gkotsinas2022}, would also correct the blind spot we currently have in our understanding of the evolution from the outer solar system to the JFC population.


\section{Summary}\label{sec:summary}

We investigate the evolution of pits on the surface of JFCs visited by space missions, i.e. 81P/Wild~2, 9P/Tempel~1 and 103P/Hartley~2, by applying the same method as for 67P/Churyumov-Gerasimenko \citep{benseguane2022}. On each comet shape model, we select facets to sample at least 10 pits across the surface, distributed at all latitudes. The energy balance at the surface is then computed by including shadowing and self-heating contributions and used as a boundary condition of a 1D thermal evolution model to quantify the amount of erosion sustained after a number orbital revolutions. 
This number is selected for each comet to correspond to the number of perihelion passages on the current orbit: 6 orbits for 81P, 13 orbits for 9P and 20 orbits for 103P. We find that:
\begin{itemize}
    \item[1-] Similarly to what was found for 67P \citep{benseguane2022}, erosion resulting from water-driven activity is primarily controlled by direct insolation. Strong seasonal patterns thus arise. However, our results suggest that erosion depends not only on the peak and total energy the surface receives, but also on how energy is delivered as a function of time, to produce subtle effects that appear unique to each comet nucleus, and in particular its shape and rotational properties.
    
    \item[2-] Progressive erosion sustained after multiple perihelion passages is not able to carve large depressions of the observed size and shape on any of the comets we studied.
    
    \item[3-] Cometary activity tends to erase sharp morphological features: they become wider and shallower over time.
    
    \item[4-] Because the same patterns hold for four comet nuclei, our results can reinforce the evolutionary sequence evidenced from independent measurables such as surface topography and roughness \citep{vincent2017a}, the phase function coefficient \citep{kokotanekova2018}, or the rate of observed mini-outbursts \citep{kelley2021}, that transforms ``young'' cometary surfaces, with sharp surface topography prone to outbursts into ``old'' cometary surfaces.

    \item[5-] We suggest that the mechanism at the origin of pits on JFCs should be able to carve these features in a region of the solar system where water ice does not sublimate: the Centaur phase of JFCs thus appears critical to understand their surface properties. 
\end{itemize}

\begin{acknowledgments}
We warmly thank Marc Costa Sitja, Alfredo Escalante Lopez, Dave Schleicher, as well as Boris Semenov, for their assistance with SPICE kernels, especially relating to 103P/Hartley~2. We also thank the anonymous referee whose comments helped improve this manuscript. 
This work has benefited from fruitful discussions supported by the International Space Science Institute in the framework of International Team 504 ``The Life Cycle of Comets'' led by Rosita Kokotanekova.
This project has received funding from the European Research Council (ERC) under the European Union’s Horizon 2020 research and innovation programme (Grant Agreement No 802699).
We gratefully acknowledge support from the PSMN (Pôle Scientifique de Modélisation Numérique) of the ENS de Lyon for the computing resources. AB was supported by grants ST/S000364/1 and ST/W001071/1. 
\end{acknowledgments}

\bibliography{pits}{}
\bibliographystyle{aasjournal}



\end{document}